\documentclass{optica-article}

\makeatletter
\def\@journalname{preprint}   
\def\@articletype{}            
\makeatother

\usepackage{lineno}

\usepackage[section]{placeins} 
\begin{document}

\title{Supercontinuum Generation in 1-decanol}

\author{Nathan G. Drouillard,\authormark{1,*} Jacob A. Stephen,\authormark{1}, Chathurangani Jayalath Arachchige,\authormark{1}, Jeet Shannigrahi,\authormark{2}, and TJ Hammond\authormark{1}}

\address{\authormark{1}Dept. of Physics, University of Windsor, Windsor ON N9B 3P4\\
\authormark{2}Dept. of Physics and Astronomy, University of British Columbia, Vancouver BC V6T 1Z1, Canada\\}

\email{\authormark{*}droui116@uwindsor.ca}

\begin{abstract*} 
Although solids have been recently used in ultrafast experiments for spectral broadening due to their relatively high nonlinearity, their sensitivity to damage limits their long-term stability. Liquids are a possible alternative to solids as a nonlinear medium because of their comparable nonlinearity and resistance to permanent damage. We generate a supercontinuum in 1-decanol that spans more than an octave from the visible to the near-infrared regime. We measure the nonlinear index of refraction of 1-decanol and find a significant $n_4$ contribution. This contribution leads to a nonlinearity comparable to CS$_2$ (a frequent reference for nonlinear optics) in high-intensity regimes while being significantly less volatile and toxic. We find this supercontinuum spectrum to be stable for at least 30 minutes.
\end{abstract*}

\section{Introduction}

Ultrashort and few-cycle laser pulses require supercontinuum (SC) spectra spanning more than an octave of bandwidth \cite{Liu_2012}. Few-cycle laser pulses offer a route of coherent control on a femtosecond (1 fs $= 1\times 10^{-15}$s) timescale, driving experiments in strong-field physics, high-harmonic generation, and attosecond science \cite{Gaumnitz_2017,Krausz_2009,Hammond_2017,Goulielmakis_2008,Hassan_2016}. SC generation has been successfully achieved in gases \cite{Gao_2022,Adamu_2019,Kim_2008} and in solids \cite{Heidt_2011,Grigutis_2020,Mar_iulionyt__2021} with spectra spanning more than an octave. More recently, SC generation in liquids has been a growing area of study, particularly in hollow-core fibres \cite{Chu_Van_2021} and liquid-core photonic-crystal fibres (PCFs) \cite{Thi_2022}. 

Liquids are favourable media for nonlinear optics due to their high density, high nonlinearity, and good power stability \cite{Schaarschmidt}. Ethanol has successfully generated SC in a PCF \cite{Kumar_2020}, although its volatility and low boiling point limit its stability \cite{ethanol}. Among the commonly used and effective liquids for SC generation are chloroform (CHCl$_3$)\cite{Van_Lanh_2019}, carbon disulfide (CS$_2$)\cite{Wang_2019}, toluene (C$_7$H$_8$)\cite{Thi_2022}, and benzene (C$_6$H$_6$)\cite{Chu_Van_2021}. A significant drawback of using these four liquids is that they are all highly toxic \cite{Lionte_2010,Yan_2019,Cruz_2014,Snyder_1993}. Due to the volatility and associated health risks with these organic solvents, we began studying high boiling point organic alcohols instead. The MAK (“maximale Arbeitsplatz-Konzentration”: maximum workplace concentration) \cite{Forschungsgemeinschaft} of 1-decanol is reportedly 10 ~mL/m$^3$ with no reported carcinogenicity \cite{Hartwig}, compared to a value of 5 ~mL/m$^3$ for carbon disulfide \cite{Stevenz}, 0.5 ~mL/m$^3$ for chloroform \cite{Chloro}, 5 ~mL/m$^3$ for benzene \cite{Angerer}, and 50 ~mL/m$^3$ for toluene \cite{Tol} (but toluene vapours are known to be harmful in comparison to 1-decanol \cite{noaa}).

We have previously studied the effects of 1-decanol on intense femtosecond pulses and found that its high boiling point (231.85~$\,^{\circ}\mathrm{C}$)\cite{1-decanol} leads to stable pulse compression by a factor of three that is readily achievable \cite{Stephen_2022}. In this paper, we further explore the nonlinear optical properties of 1-decanol and demonstrate significant contributions from higher-order nonlinearities. 

\section{Methods}

As shown in Figure 1(a), we use a Ti:Sapphire laser that produces 100$~$fs, 1.5$~$mJ pulses centered at 785$~$nm with 1 kHz repetition rate. The beam passes through a half-wave plate ($\lambda/2$) and a polarizer (Pol.) to control the power. A 200$~$mm focal length lens focuses the beam into the 1$~$cm cuvette of 1-decanol, which sits on an automated translation stage (Thorlabs PT1-Z8). When measuring the spectrum as a function of axial position, the output beam is then focused into the spectrometer with a 100$~$cm focal length lens and attenuated before reaching the spectrometer (Ocean Insight Flame-S) using neutral density filters. When measuring the spectrum as a function of radial position, the spectrometer is placed 45$~$cm after the 200$~$cm focusing lens with no second lens after the cuvette so the beam is not focused into the slit of the spectrometer; instead, the spectrometer is scanned through the beam radially. 

\begin{figure}[!htbp]
  \centering
  \includegraphics[width=0.75\columnwidth]{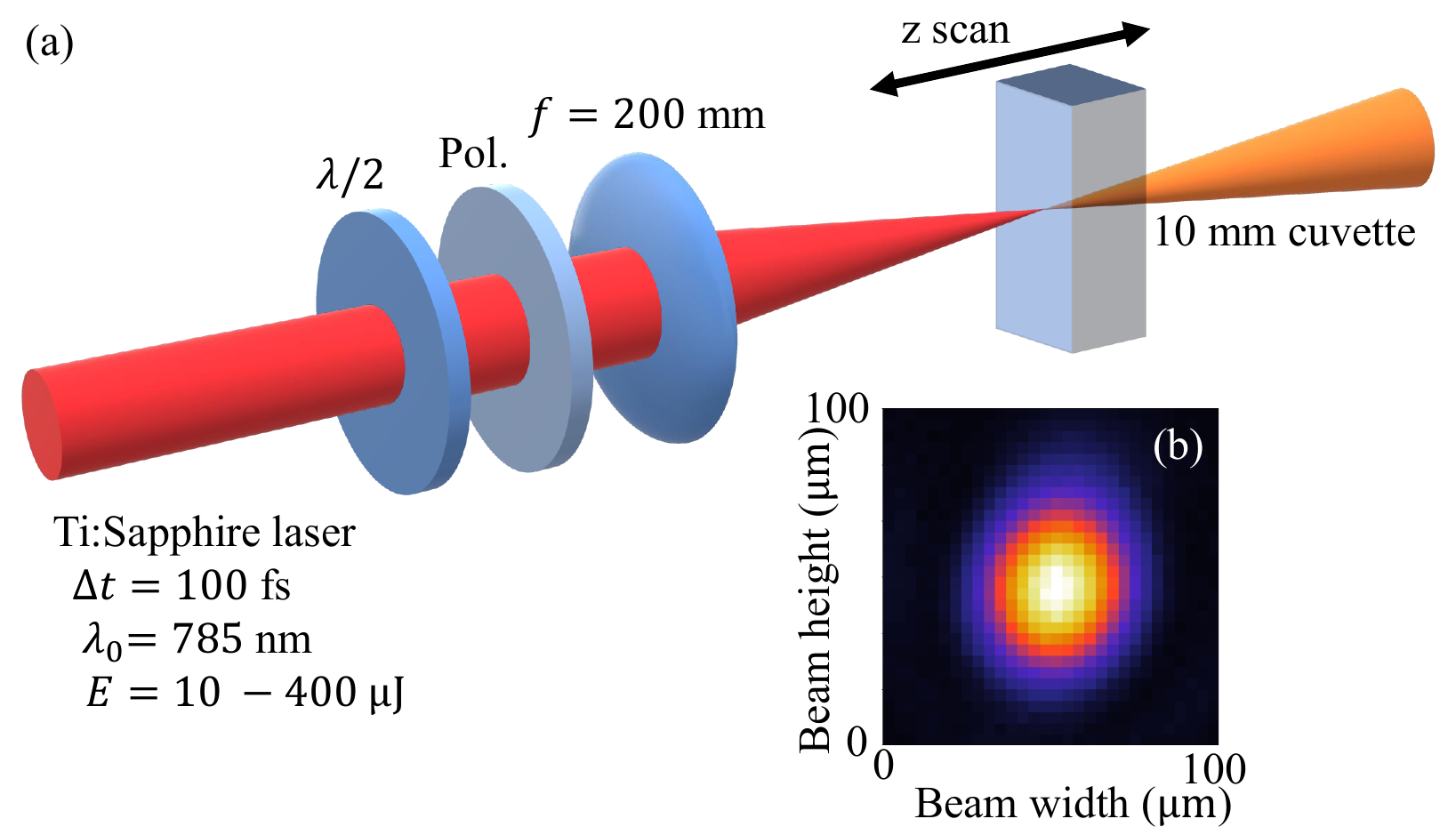}
  \caption{(a) Experimental setup used for supercontinuum generation in 1-decanol. The beam travels from left to right passing through a half-wave plate ($\lambda/2$), followed by a polarizer (Pol.), a 200$~$mm focal length lens, and finally into a cuvette which sits on an automated translation stage. Not shown is the spectrometer that measures the spectrum of the output beam. (b) An image of the radial beam profile is shown. The beam waist is measured to be 26~\textmu m at the focus.}
\end{figure}
\FloatBarrier

In both the radial and axial cases, the data is taken using 3 wavelength filters and stitched together during data analysis to increase the dynamic range and to better resolve both the visible and infrared portions, which are relatively weaker in intensity compared to the central wavelength. To this end, the spectrum is measured with no filter in order to resolve the spectrum from 740-800$~$nm. Next, the same measurement is made using a visible filter (335-610$~$nm), and two infrared (IR) filters (800-850$~$nm and 850$~$nm long-pass). Normalization is required to account for different spectrometer integration times and different degrees of beam attenuation before the spectrometer.

\section{Results and Discussion}
\subsection{Measuring the nonlinear index of refraction}

\begin{figure}[!htbp]
  \centering
  \includegraphics[width=0.75\columnwidth]{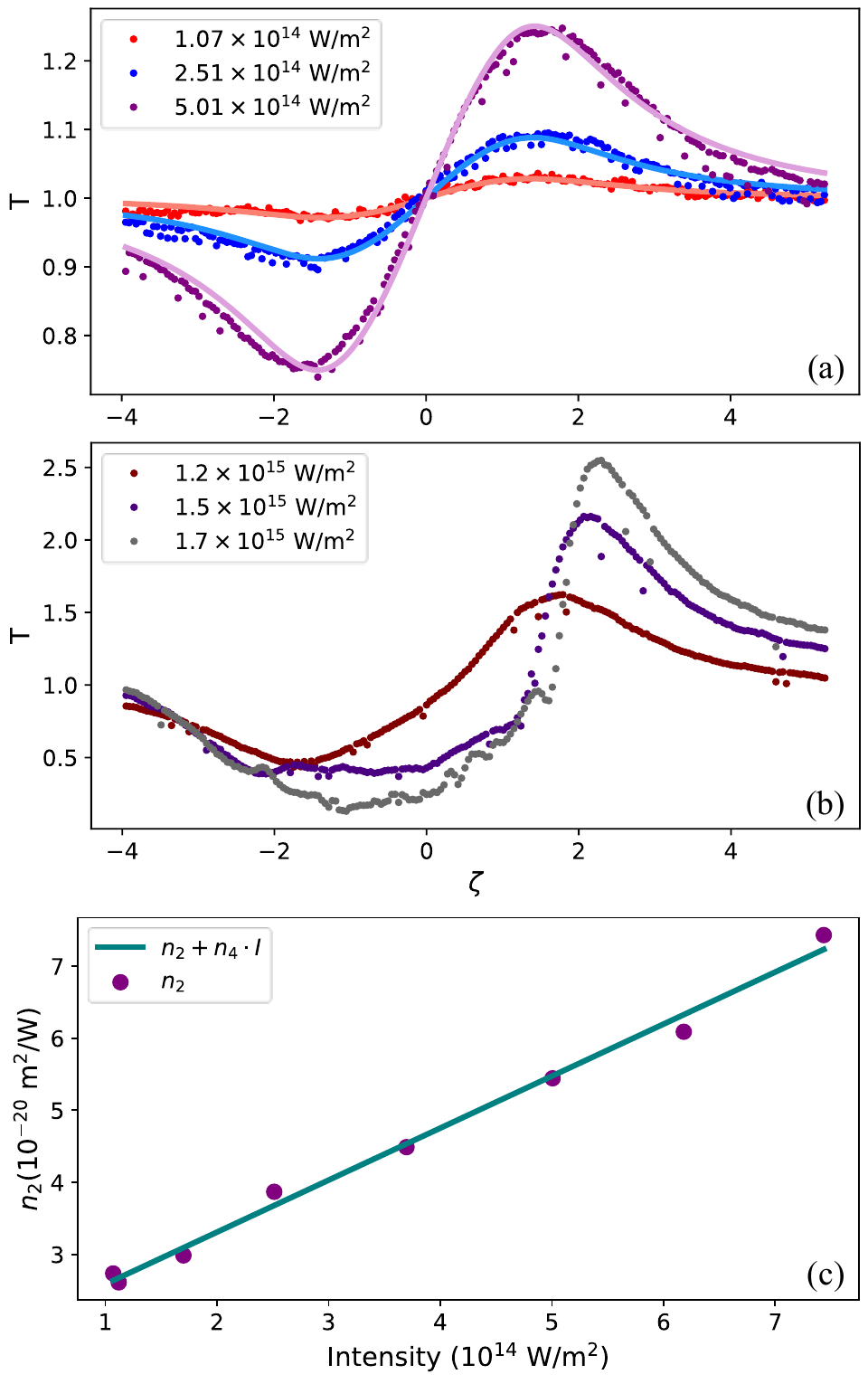}
  \caption{(a) Low-intensity regime. The points are measured data while the solid lines are the corresponding curve fits. T is the transmittance, which is a function of normalized length $\zeta=\frac{z}{z_0}$ where $z$ is the direction of beam propagation and $z_0$ is the Rayleigh range\cite{Stephen_2022}. (b) High-intensity regime, showing deviation from the z-scan fit. (c) Plotting $n_2(I)$ in order to calculate the fifth-order nonlinear index of refraction,$n_4$.}
\end{figure}
\FloatBarrier

We measure the nonlinear index of refraction of 1-decanol with the z-scan technique \cite{KuzukandDirk}. We use an index of refraction of 1.43\cite{Ortega_1982}, which agrees with our previous measurement of the dispersion and index of refraction of 1-decanol\cite{hammond}. Figure 2(a) shows the low-intensity regime where we do not see SC, but the data follows a typical z-scan shape. For this reason, the curve fit agrees extremely well with the data, allowing us to calculate $n_2 = 1.87\times10^{-20}$~m$^2$/W at I$_0$ = 0. Aside from our previous measurement \cite{Stephen_2022}, the only known value for the n$_2$ of 1-decanol is $10.9\times10^{-20}$~m$^2$/W, which was measured using a 1.06~\textmu m ps laser \cite{Ho_1979}. It should be noted that the nonlinear phase imparted by the cuvette is small compared to that of the liquid, due to the thickness of the cuvette walls being much smaller than that of the liquid. Nevertheless, we do include the nonlinearity of the cuvette in our model \cite{Milam}.

We record the closed-aperture transmission and fit it to the function, 
\begin{align}
T = 1 + \Delta \Phi F(\zeta, l)
\end{align}
where $\Delta \Phi = (2\pi/\lambda) n_2 z_0 I_0$ is the self-focusing-induced phase shift, $\lambda$ is the laser central wavelength, $z_0$ is the (free space) Rayleigh range, and $I_0$ is the peak intensity. The characteristic z-scan function for a thick sample,
\begin{align}
F(\zeta,l) = \frac{1}{4} \ln \left( \frac{\left[(\eta + l/2)^2 + 1 \right] \left[(\eta - l/2)^2 + 9 \right]}{\left[(\eta - l/2)^2 + 1 \right] \left[(\eta + l/2)^2 + 9 \right]} \right),
\end{align}
where the parameters $\zeta = z/z_0$ and $l = L/z_0$ scale the scanning length and the cuvette length, respectively. The material nonlinear Kerr coefficient, $n_2$, is then extracted from the fit from the z-scan curves.

Figure 2(b) shows the high-intensity, SC regime where the data do not follow the z-scan shape exhibited in (a). Because the z-scan technique breaks down in this regime, we do not use it to calculate higher orders of nonlinearity for the medium. The breakdown of the fit begins at $>1\times10^{15}$~W/m$^2$ and becomes more apparent $>2\times10^{15}$~W/m$^2$. We argue that multi-photon absorption is not responsible for the disagreement with the z-scan fit, because we do not measure additional power loss through the cuvette at these high intensities.

Figure 2(c) illustrates how in the low-intensity regime, we are able to fit the nonlinear index as a function of intensity in order to calculate $n_4 = 7.22\times10^{-35}$~m$^4$/W$^2$. Compared to some values of $n_2$ reported in the literature for CS$_2$, a commonly used medium for supercontinuum generation, the $n_2$ of 1-decanol is lower by approximately an order of magnitude ($n_2$(CS$_2$)$\approx2.8\times10^{-19}$~m$^2$/W)\cite{Couris_2003,Ganeev_2006}. However, $n_4$ for CS$_2$ is reportedly $n_4 = -2\times10^{-35}$~m$^4$/W$^2$ at 800$~$nm \cite{Kong_2009}, where the change of sign is related to the location of the third harmonic compared to the absorption bands of the medium \cite{Besse_2014}. Therefore, we have shown that 1-decanol exhibits a degree of optical nonlinearity on par with CS$_2$ without a high degree of toxicity. In fact, using the cited values, the nonlinear index of refraction of 1-decanol exceeds that of CS$_2$ above $2.8\times10^{15}$~W/m$^2$, which is the regime where we report SC generation. 

\subsection{Spectrum vs. axial position}

\begin{figure}[!htbp]
  \centering
  \includegraphics[width=0.75\columnwidth]{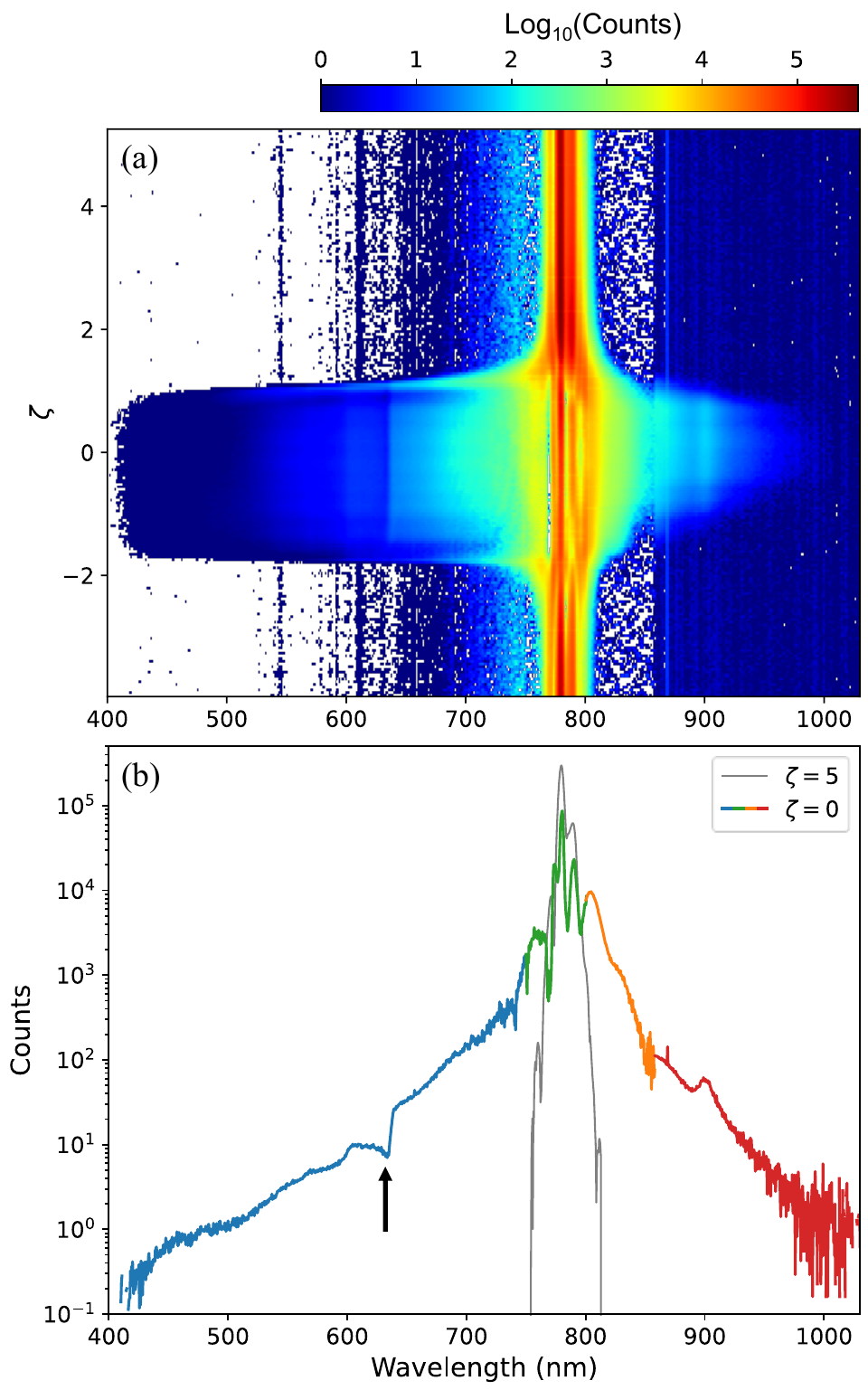}
  \caption{(a) As the cuvette passes through the focus ($\zeta = 0$~mm), the spectrum rapidly broadens and comprises frequencies across more than an octave. (b) A lineout of the spectrum at the focus ($\zeta=0$) and far from the focus ($\zeta=5$). Colours indicate the bandwidth of each optical filter used in the measurement; arrow indicates position of the Raman loss.}
\end{figure}
\FloatBarrier

Figure 3(a) illustrates the intensity of the spectrum as a function of wavelength and cuvette position. As the 1-decanol approaches the focus, the spectrum undergoes significant spectral broadening, creating a SC which is stable until the 1-decanol passes out of the focus and returns to its initial bandwidth. The SC contains frequencies across more than an octave, from approximately 450$~$nm to$~$950 nm. This sudden explosion in spectrum near $\zeta = -2$ indicates that the beam undergoes significant self-focusing, which extends beyond the Rayleigh range and indicates filamentation. At these intensities, the interplay between absorption, thermal lensing, self-focusing, and diffraction allows for extending the high-intensity interaction length, enabling more efficient SC generation \cite{Falcao_PRL_2013, Bautista_JOSAB_2021}.

Figure 3(b) shows lineouts contrasting the initial spectrum ($\zeta = 5$) to the SC ($\zeta = 0$). The peak intensity of the beam at the focus is estimated to be $3\times10^{15}$~W/m$^2$. The difference in energy between 785$~$nm and 636$~$nm corresponds to a Raman shift of -2984$~$cm$^{-1}$, which lies within the broad Raman response of 1-decanol centered at 2900$~$cm$^{-1}$ \cite{Carrizo_2020}, indicating Raman loss caused by the pump (black arrow) \cite{Yang_2007,Santhosh_2010,Sreeja_2013, DrouillardAppSpec2025}. 

\subsection{Spectrum vs. radial angle}

The radial distribution of the spectrum, as shown in Figure 4, is such that all wavelengths in the supercontinuum are present near the center of the beam. However, while the visible wavelengths remain present at greater radial distances, the infrared components do not. Furthermore, we do not observe significant spectral peak shifts off-axis.

The visible portion of the spectrum is relatively uniform and not significantly radially dependent over this divergence. This angle-wavelength distribution exhibits a different pattern than the X-wave amplification exhibited from four-wave mixing \cite{Faccio_PRL_2006}. This pattern shows that the beam maintains a tight focus for an extended length, indicating that the increased spectral broadening comes from filamentation.

\begin{figure}[!htbp]
  \centering
  \includegraphics[width=0.95\columnwidth]{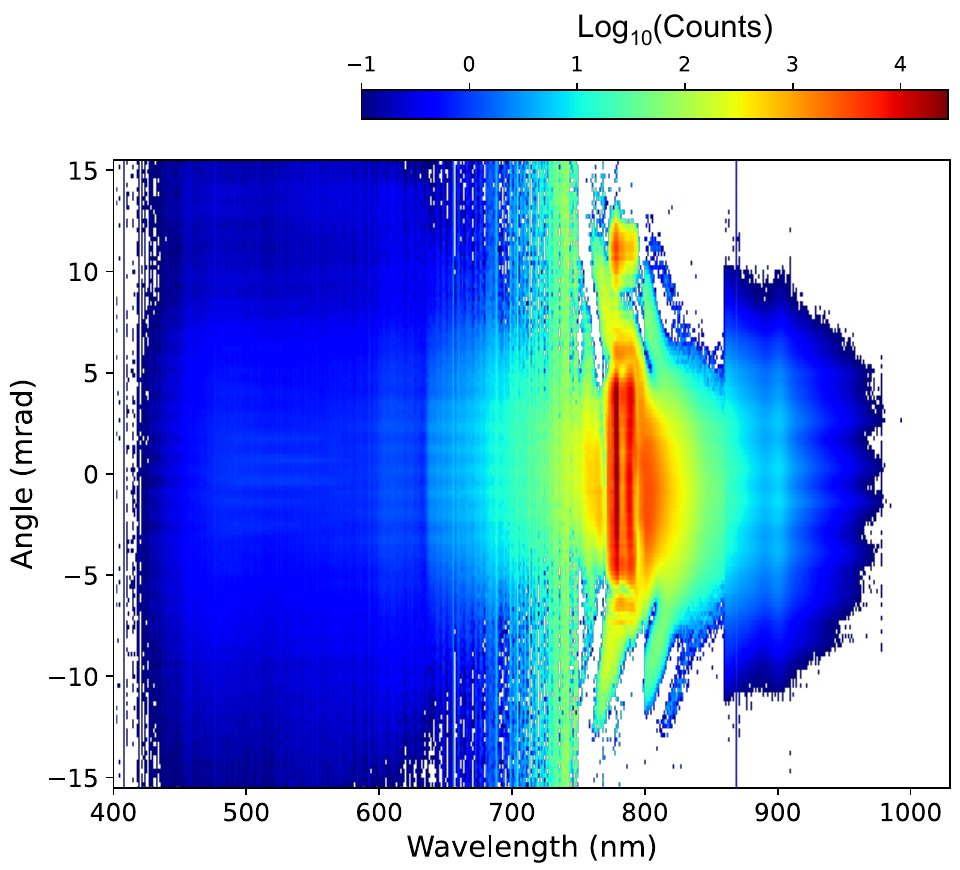}
  \caption{The broad spectrum produced in 1-decanol as a function of radial angle from the center of the output beam. The visible portion exhibits a large divergence, indicative of the tight spatial confinement in the supercontinuum generation process.}
\end{figure}
\FloatBarrier

\subsection{Temporal stability of the spectrum}

\begin{figure}[!htbp]
  \centering
  \includegraphics[width=0.75\columnwidth]{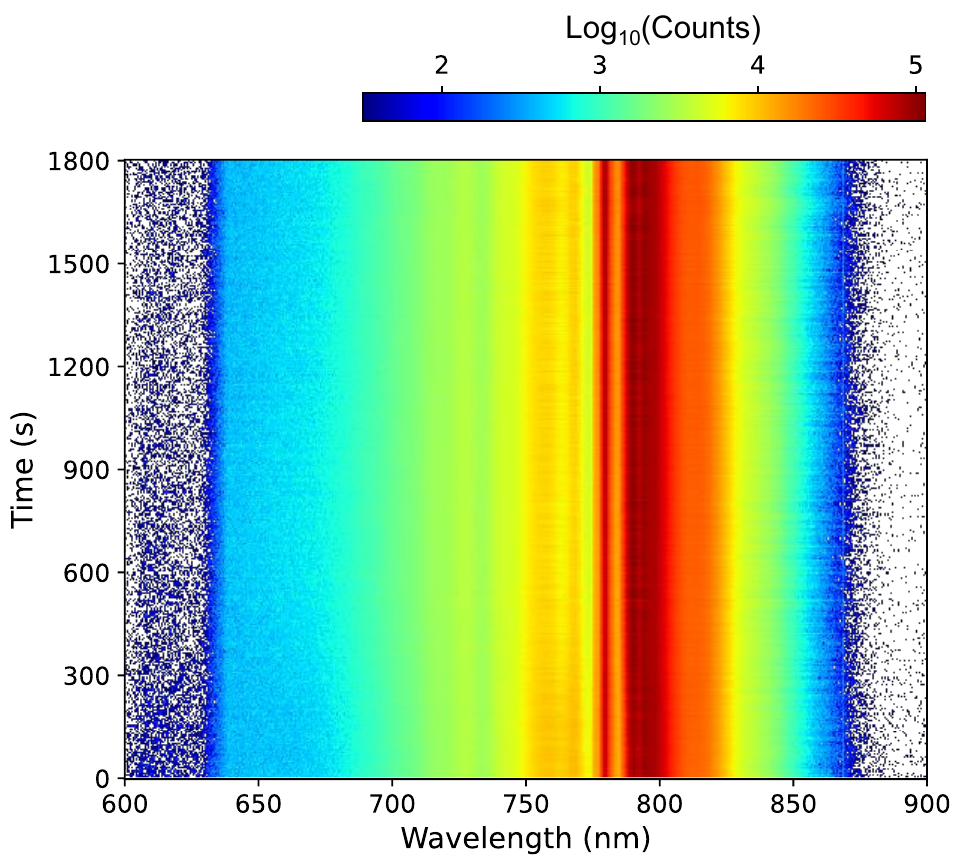}
  \caption{The temporal stability of the most intense portion of the spectrum over 30 minutes, demonstrating little variation in the supercontinuum.}
\end{figure}
\FloatBarrier

Laser power fluctuations and material degradation would cause instability in the edges of the supercontinuum. As shown in Figure 5, we measured the temporal stability of the SC over the main portion of the spectrum for 30 minutes. A benchmark of stability for SC spectra in liquids is found in \cite{Schaarschmidt} where the authors report a spectrum stable for over 70 hours using CS$_2$ in a liquid-core fiber.  Because of the high boiling point of 1-decanol, we observe little evaporation of the sample after several weeks of repeated use and we expect similar longevity in our sample.

\section{Conclusion}
We show that it is possible to generate a supercontinuum in 1-decanol that contains frequencies across more than an octave. Although the nonlinear Kerr coefficient is modest, we find that the higher-order nonlinearities significantly contribute to the measured phase shift. Not only is 1-decanol a much safer alternative to commonly used liquids such as chloroform (CHCl$_3$), carbon disulfide (CS$_2$), toluene (C$_7$H$_8$), and benzene (C$_6$H$_6$),  but it demonstrates a nonlinear index on par with these liquids at high intensities. Additionally, we find that the supercontinuum is stable for extended periods of time, enabling prolonged study of strong-field physics. We expect that the comparable nonlinearity of 1-decanol to more toxic liquids makes it an attractive alternative for future nonlinear optics applications such as Kerr-instability amplification. 

\subsection{Data Availability}

\begin{enumerate}
\item TJ Hammond, "Data to replicate supercontinuum generation in 1-decanol," figshare (2014), \url{https://doi.org/10.5683/SP3/4HLCO9}.   
\end{enumerate}

\section{Back matter}

\begin{backmatter}
\bmsection{Funding}
N.G. Drouillard acknowledges funding from the Queen Elizabeth II Graduate Scholarship in Science and Technology (QEII-GSST) Program, sponsored by the Ontario government. Jeet Shannigrahi acknowledges funding from Mitacs. TJ Hammond acknowledges funding from NSERC Discovery RGPIN-2019-06877 and University of Windsor Xcellerate grants.

\bmsection{Acknowledgment}
We want to thank Aldo DiCarlo for his technical assistance and Jill Crossman for her support.  

\bmsection{Disclosures}
``The authors declare no conflicts of interest.''

\bmsection{Data Availability Statement}

\bmsection{Data availability} Data availability. Data underlying the results presented in this paper are available in Ref. \cite{Data}.

\bigskip

\end{backmatter}

\section{References}

\bibliography{bibliog}

\end{document}